\documentclass[10pt]{article}
\title{\it A Tribute to Alain Colmerauer}
\author{
\small{{JACQUES COHEN}}\\ 
\small{\textit{TJX/Feldberg Professor}}\\
\small{\textit{Volen Center}}\\ 
\small{\textit{Brandeis University}}\\ 
\small{\textit{Waltham, MA 02454, USA}}\\ 
\small{(\textit{e-mail:}\ \texttt{jc@cs.brandeis.edu})}}
\date{July 10, 2001}

\begin{document}
\maketitle

\section*{\centerline{Prelude}}
As an invited contributor to this Festschrift honoring Alain 
Colmerauer, I feel compelled to give not only an account of his 
main research contributions, but also of my perspective on the 
motivations behind them. I hope that this will provide the reader 
with a glimpse of how a focused, tenacious, rigorous, and inventive 
mind like Alain's picks research problems and proceeds to solve 
them. 
The history of Prolog, the language that remains one of Alain's 
major accomplishments, is well documented. His paper on the 
\textit{``Birth 
of Prolog,''} co-authored with Philippe Roussel \cite{CR96}, is a 
highly recommended account of the circumstances that led to the 
development of Prolog. Bob Kowalski \cite{Kow88} presents his views 
of the early history of Prolog from the automatic theorem proving 
perspective. Finally, my own paper on the topic \cite{Coh88} contains 
material complementing Alain's and Bob's narratives. 
Instead of recasting already-available historical material, I 
have opted to present here a more personal account of Alain's 
contributions, acknowledging in advance the individual bias inherent 
in such an accounting of long-past events.

\section*{\centerline{An Early Encounter: the Sixties}}
I have been fortunate enough to work closely with Alain Colmerauer 
for almost four decades. We met in the early fall of 1963, when 
Alain was in his early twenties, had just completed his undergraduate 
studies at the Institut Polytechnique de Grenoble, France, and 
was contemplating a doctoral degree.
I had been attracted to Grenoble by the expansion that was taking 
place in the development of the new field of computer science, 
an expansion based on applied mathematics. At that time, the 
education in applied mathematics emphasized mostly numerical 
analysis and Boolean algebra. The Institute of Applied Mathematics 
in Grenoble (known by its French acronym IMAG) was led by Professor 
Jean Kuntzmann, a specialist in Boolean algebra; his closest 
associate was Professor Noel Gastinel, an expert in numerical 
analysis. 
In addition to Professors Kuntzmann and Gastinel, two younger 
researchers were prominent in the faculty at the Institute: Louis 
Bolliet, an experienced programmer of the earlier computers, 
and Bernard Vauquois, an astronomer by training, who became involved 
in formal language and automata theory and its application to 
automatic natural language translation.
There were two events that made 1960s Grenoble an exciting place 
for research in computer science. The first was an on-going effort 
by European and American researchers to design and implement 
a standard computer language, Algol60, based on the experience 
gained in developing earlier languages such as Fortran. The second 
was the availability of a mainframe computer (IBM 7044) that 
offered the IMAG researchers superior opportunities for computation 
in those early days.
Most of what we learned in Grenoble at that time consisted of 
novel techniques that had been recently proposed and published 
in \textit{Communications} and the \textit{Journal} of the ACM, or in 
contemporary 
monographs and dissertations. I recall that Bernard Vauquois, 
a member of the Algol60 original design team, and who was then 
directing a group on mechanical translation, was teaching a course 
in languages and automata theory. The course was purely theoretical, 
and it was our responsibility to work toward application of those 
theories to language implementation using the existing computers.
One of the main goals of the computer science team at IMAG was 
to develop an Algol60 compiler for the newly acquired IBM machine. 
Jean-Claude Boussard was the doctoral student responsible for 
developing the compiler. This software would be among the first 
Algol compilers developed in France.
Algol60 had a rigorous definition of its syntax, using what is 
known today as Backus-Naur-Form or BNF. In those days syntax-directed 
compilers were studied at the research level, but were considered 
inefficient for implementation in the available machines. Nevertheless, 
the graduate students in the compiler group of Louis Bolliet, 
including Alain and myself, were fascinated by the possibility 
of using syntactic rules as templates for building compilers.

\section*{\centerline{Alain's First Research Project}}
The project that would tentatively make up Alain's dissertation 
was to design and implement an error-detecting program to be 
used as a front-end for a Cobol compiler. Cobol's syntax was 
available in BNF, but Alain's goal was to design a general syntax-directed 
general program. The implementation itself was to be written 
in Algol60 using the compiler being developed in Grenoble.
At that time, there were very few papers available on compilers. 
The prevalent approach was that of Dijkstra's stack-machine model 
which eventually became available in the book of Randell and 
Russell \cite{RR64}. Through Louis Bolliet we obtained an interesting 
new paper by Robert W. Floyd, who was at the time working at 
Computer Associates, near Boston. The paper was entitled ``Syntactic 
Analysis and Operator Precedence,'' and had appeared in a recent 
(1963) issue of the \textit{Journal} of the ACM \cite{Flo63}. Basically, 
Floyd had found a way of automatically generating Dijkstra's 
stack operator precedences for languages exhibiting a special 
restricted form of grammar rules.

I recall that Alain became extremely interested in that paper 
and decided to use Floyd's precedence grammars in his syntax-driven 
error-detector for Cobol. This was easier said than done. Anyone 
familiar with that approach realizes that Cobol's syntax does 
not easily conform to a precedence form. There were a multitude 
of precedence conflicts that would have to be resolved ``manually,'' 
that is, case by case. In addition, precedence grammars, being 
deterministic, would not allow rules with identical right-hand-sides 
and that characteristic was common in the existing BNF definitions 
for various languages.

Thus, Alain's initial problem was more difficult than he had 
anticipated. The reader will soon realize that what happened 
in bypassing that problem is typical of Alain's reaction when 
confronted with an obstacle. 
I open a parenthesis to mention a couple of Alain's traits that 
shed some light onto his creativity and perseverance. We used 
to drive through the narrow old streets of Grenoble and surrounding 
towns. Alain seldom took the same route twice: his innate curiosity 
often led him to find new ways of going from one spot to another. 
That temperament obviously served him well when it came to problem 
solving. I also remember that Alain had rented a studio apartment 
in one of the boulevards of Grenoble. He loved and still loves 
sports, sailing being one of his favorite hobbies. I recall that 
he had decided to build a small sailing boat in his studio; when 
the boat was ready, colleagues and I helped Alain remove it through 
a window. I am sure that he had taken the necessary precautions 
by careful measurements of the room, prior to undertaking the 
unusual project. Again it seemed that Alain had the knack for 
generating clever problems and then surmounting them. (Perhaps 
even his choice of living in a street named \textit{Impasse des Iris} 
is not completely random!) 
Now, to continue with the precedence grammar problem to which 
Alain had applied himself. A paper of Griffiths and Petrick also 
came to our attention \cite{GP65}. It involved the design of a two-stack 
Turing Machine (TM) to estimate the efficiency of various context-free 
parsing methods, in particular the approaches known as top-down 
and bottom-up. The authors had cleverly simulated various existing 
parsers using TM sets of instructions. 
I remember Alain avidly reading that paper. The notion of nondeterminism 
was implicit in the TM instructions; the efficiency of various 
parsers was estimated by simulating the TM in a computer and 
by determining the number of steps needed to parse representative 
strings generated by typical grammars.
Alain's acumen in bypassing difficulties with precedence conflicts 
amounted to generalizing Floyd's precedence parsers to be able 
to process more general languages than those advocated by Floyd. 
Basically, bottom-up and shift-reduce parsers replace the right-hand-side 
of a rule by its left-hand-side. When parsing from left to right, 
the element on the top of the stack is compared with the current 
element in the string being parsed. That introduces asymmetry, 
as only the stack may contain non-terminals allowing the parser 
to manipulate them.
By using two stacks, \textit{\`{a} la} Griffiths and Petrick, symmetry 
is restored since the input string being parsed is placed in 
a second stack, and reductions may occur in either stack. This 
extension allows for the parsing of a significantly more general 
class of languages than those defined by simple precedence. It 
then became possible to handle parsing and error detection based 
on the existing Cobol syntactic rules, virtually without ``manual'' 
intervention.

\section*{\centerline{A Premonition for Prolog: the Late Sixties }}
Alain's JACM paper on Total Precedence summarizes his dissertation 
and provides a preview of the ingeniousness, simplicity and rigor 
he applied to the solving of a fairly complex computer science 
problem \cite{Col70}. The dissertation can also be viewed as containing 
ingredients that appeared later on in the development of Prolog 
(e.g., parsing and nondeterminism).
The language Algol68 was being perfected at about the time of 
the completion of Alain's thesis. He showed great interest in 
the two-level grammars proposed by van Wijngaarden to define 
the syntax of that new language \cite{Wij68}. Again, that formalism 
had some intriguing resemblance with the one that later became 
Prolog rules. (This is because two-level grammars can represent 
a potentially infinite number of context-free rules.)
Around 1967, Grenoble's compiler team familiarized itself with 
yet another paper by Robert Floyd; in this paper, he proposed 
annotations to a computer language that allowed its processor 
to deal with (\textit{don't know}) nondeterministic situations 
\cite{Flo67}. 
Floyd also proposed an implementation of his ideas using 
flowcharts. A group of graduate students including Alain actually 
implemented a version of nondeterministic Algol60 that proved 
successful in describing succinctly the solution of combinatorial 
problems.
I offer the above reminiscences of the precursors to Prolog because 
I firmly believe that the papers of Floyd, Griffiths \& Petrick, 
and van Wijngaarden were pivotal in establishing a frame of mind 
that prepared Alain for the ``discovery'' of Prolog. 
If I recall correctly, in one of the Algol68 design meetings, 
Alain had suggested to van Wijngaarden the incorporation of 
nondeterministic 
constructs into Algo68; the latter replied with something like: 
``Wait young man, one should not introduce a feature into a language 
just because of it being nifty.'' It seems that it behooved to 
Alain to do precisely that a decade later.

\section*{\centerline{The Stay in Montreal}}
Around 1967, upon finishing his doctoral degree at Grenoble, 
Alain spent three years at the University of Montreal. The University 
of Montreal and other Canadian universities were attractive options 
for the military service of young Frenchmen, including Alain. 
While in Montreal, Alain decided to concentrate his research 
on natural language processing and artificial intelligence (AI). 
His interactions were with both computer scientists and linguists, 
and his decision to include both disciplines in his research 
must have been due in no small part to his wife, Colette, who 
is an accomplished linguist.
During that period Alain developed Q-systems, now considered 
the precursor of Prolog's operational semantics. Essentially 
it consists of a set of rules specifying that a sequence of trees 
can be rewritten into another sequence of trees; a version of 
that model was subsequently used by Alain to define rigorously 
the semantics of later versions of Prolog.
In my view, Alain was continuing to generalize his work on context-free 
parsers to include nondeterminism. That is a key feature in 
natural language processing, where dealing with ambiguities is 
a must. In addition, since syntax \textit{per se} is insufficient to 
deal with semantics, Alain embarked on an in-depth study of 
theorem-proving 
techniques and became aware of the famous paper by Alan 
Robinson on resolution and unification. That paper had been published 
two years earlier \cite{Rob65}. It was Cordell Green in 1969 who 
had proposed using theorem-proving as an approach to problem 
solving \cite{Gre71}.
When Alain was considering returning to France in 1969 he had 
multiple choices. Under the sponsorship of Robert Floyd, Alain 
had an interview for an appointment at Stanford University. The 
final choice of Marseille is typical of what one would expect 
from Alain. He could easily have had a position in a French university 
with an already-established group in computer science, but he 
preferred to start his own department from scratch. He must also 
have been fascinated by the natural beauty of the neighboring 
towns in Provence, like Aix-en-Provence and Cassis. 
As a good hiker and sailor, 
the calanques (fjord-like inlets) in Cassis must also have exerted 
a strong attraction.

\section*{\centerline{Back to France and the Dawn of Prolog: the Seventies}}
Upon returning to France and settling in Marseilles, Alain had 
the challenging task of starting a new computer science department 
at the campus of Luminy. He surrounded himself by bright students, 
among them Philippe Roussel, and concentrated his research on 
theorem-proving and computational linguistics. The computing 
facilities in Marseilles were minimal, and this must have been 
anticlimactic considering the good equipment available in Grenoble 
and Montreal. The struggle to obtain adequate computers is one 
that Alain had to face throughout his tenure as chair of the 
budding CS department in Luminy.
In the late sixties and early seventies the artificial intelligence 
group at Edinburgh was among the best in Europe. A team there 
was exploring the potential of automatic theorem-proving techniques 
to problem solving. A doctoral student from Edinburgh, Bob Kowalski, 
had shown how to reduce substantially the search space for 
resolution-based 
theorem-provers \cite{KK71}. Alain obtained funds to invite Bob for 
a stay in Marseilles and the cooperation between Edinburgh and 
Marseilles flourished. As I mentioned earlier the history of 
that cooperation is well documented.
It is fair to state that before Alain Colmerauer's entry into 
the field of programming language design, there were basically 
two applicable paradigms: one representing imperative languages 
(like Algol or Fortran) and the other functional languages (like 
Lisp). Alain's and Bob's remarkable insight was to ``invent'' or 
``discover'' a third paradigm, known as logic programming languages 
and represented by Prolog. Prolog's simplicity and logical foundations 
contributed to its worldwide acceptance and success. Perhaps 
the definitive indication of the worldwide acceptance has been 
the adoption of Prolog as the main language for the Fifth Generation 
Japanese Computer Project. 
Alain's many contributions to the elegant usage and basic features 
of Prolog remain valid to this day. They include the widespread 
use of the so-called metamorphosis grammars or the equivalent 
definite clause grammars, the suggestion of control annotations 
(e.g., the cut), the use of lazy evaluation, and so forth. Even 
the first usage of the now ubiquitous concatenation predicate 
\textit{append} is 
due to Alain. As in the case of Lisp, that procedure allows a 
user to perform clever text processing. Furthermore, because 
of Prolog's capabilities for inverse computations, \textit{append} 
can simulate other functions including table-lookup. 
I should mention here Bob Pasero and Henri Kanoui, who were among 
Alain's first doctoral students. Bob closely examined the problems 
of natural language understanding, and Henri explored the symbolic 
formula manipulation techniques that made use of the inverse 
computation capabilities of Prolog.

\section*{\centerline{Prolog II: the Eighties}}
The above contributions, even though major, were but a prelude 
to the more ambitious design features that Alain incorporated 
into the first extension to Prolog, known as Prolog II. They 
were the unification of infinite trees and a new predicate for 
testing non-equality of those trees. Those developments took 
place in the late seventies and early eighties. It is admirable 
that Alain and his colleague Michel van Caneghem were able not 
only to design the new language features but to implement them 
in what is now recognized as a very primitive personal computer: 
an Apple II. One can only marvel that Alain and Michel had implemented 
a virtual memory system using a floppy disk in a computer with 
a tiny fast (RAM) memory! 
I recall that one of the feats that Michel and Alain incorporated 
in the Apple II- Prolog II system was the ability to abort computation 
using a \textit{control-C} command, and to manage to safeguard all 
the important information prior to issuing that command. It must 
also have been an immense source of frustration to have a virtual 
memory system based on fairly unreliable floppy disks. In any 
case, that implementation became a forerunner of what now occurs 
in a PC implementation of Prolog, with all the trimmings such 
as debugging features and garbage collection.
In the mid-eighties I had the good fortune of being invited by 
Alain to teach a compiler course at Luminy at the same time that 
John McCarthy from Stanford had been invited there to present 
seminars in non-monotonic logic. I recall with amazement the 
times in which we had the opportunity to dine together and discuss 
problems in computer science. At that time McCarthy had recently 
proposed the new area of non-monotonic logic and two of Alain's 
top students were writing their dissertations on that topic. 

I also recall that John McCarthy mentioned that he belonged to 
a futuristic society, in California, that was planning to have 
scientists spend time in a moon colony to study problem solving 
capabilities in planetary environments. He suggested that Alain 
and Colette be included among potential candidates for the lunar 
sojourn. (Knowing the adventurous side of Alain and Colette I 
am not sure that they would have completely dismissed the idea 
as farfetched.) 
In another conversation with John, Alain mentioned that in the 
late sixties he had been invited to join the Stanford faculty. 
To this John retorted: If you had accepted that offer you probably 
would not have come up with Prolog!
Let me return to the novel features of Prolog II, namely infinite 
trees and \textit{diff} (the non-equality predicate). They are 
definitely 
the precursors of constraints as understood now in the Constraint 
Logic Programming (CLP) paradigm. Behind these features is the 
desire to extend, in a clean manner, the equality and non-equality 
predicates to new domains.

\section*{\centerline{CLP and Prolog III: the Nineties}}
The addition of infinite trees and \textit{diff} enabled Alain to 
carry out yet another new design, this time introducing new data 
types and global operators. To perform equalities in linear algebra 
with the necessary rigor, one must introduce first the domain 
of rationals, and second the capability to test the satisfiability 
of systems of linear equations, inequations, and disequations. 
In Prolog III, a harmonious design included the blending of the 
domain of infinite trees with that of the rationals and also 
with two additional domains: Booleans and a new domain called 
linear lists. With Prolog III, the original Prolog becomes just 
a special case of CLP.
It is essential to point out the significant role that Alain's 
doctoral students had in the development of the later Prologs. 
The implementation of Prolog III is in itself a work of programming 
art. One had to be thoroughly familiar with the admirable abstract 
machine proposed by David H. Warren and to extend it in a substantial 
manner in order to incorporate Boolean processors \textit{\`{a} la} 
Davis-Putnam, 
simplex-like solvers capable of detecting when a variable becomes 
bound to a specific value, special garbage collectors, and so on.
All this had to be done by providing a seamless interaction among 
the four domains: infinite trees, rationals, Booleans, and linear 
lists. Furthermore, there was the implicit requirement that, 
when confronted with a standard Prolog program, the compiler 
should produce code as efficiently as a Prolog processor unencumbered 
by the new extensions. In the quest of designing the various 
components of Prolog III, Alain mentored several doctoral dissertations 
that delved deeply into the algorithmic components necessary 
to process each specific domain.
Prolog III extended Prolog's applications into the realm of numerical 
computations, which are the staple of work in linear algebra 
and in operations research (OR). After designing Prolog III, 
Alain concentrated his interest on the extremely difficult combinatorial 
algorithms needed to solve scheduling problems in OR. At that 
time Alain also became acquainted with the work of William Older 
on the incorporation of interval domains to Prolog \cite{OV90}. Alain 
saw a renewed opportunity to extend Prolog III to deal with this 
new domain. 
\section*{\centerline{Intervals and Prolog IV: the Late Nineties}}
Prolog IV can be viewed as the culmination of Alain Colmerauer's 
efforts in language design; of course, each one of Alain's major 
accomplishments was considered as the culmination of the previous 
one! Prolog IV's balanced design surpasses that of Prolog III. 
The introduction of interval variables not only enables a machine-oriented 
rigorous definition of reals (the floating-point numbers), but 
it also subsumes rationals, integers, and Booleans. Essentially, 
Booleans are a particular case of finite domains, and the latter 
are a particular case of real variables expressed in floating-point 
notation. The introduction of interval variables also paved the 
way for dealing with non-linear numerical problems and enabled 
the practical solution of scheduling problems that previously 
required hours of computations.
In addition, interval variables allow for proving propositions 
asserting the nonexistence of solutions to systems of equations 
or inequations in which variables are required to have values 
within certain ranges. When confronted with such situations, 
if a processor for an interval constraint language replies ``no,'' 
then it implicitly provides a proof that no solution exists (i.e., 
assuming that the interpreter is proven correct).

\section*{\centerline{A Continued Love for Puzzles}}
Alain has always had an interest in solving mathematical puzzles. 
In the realm of Booleans, Alain admired the logical puzzles of 
Lewis Carroll (and not incidentally, Alain's youngest daughter 
was named Alice). He continually seeks out new puzzles in the 
French daily \textit{Le Monde} and in \textit{Scientific American}. 

In this context, one of the latest combinatorial problems holding 
Alain's interest is finding the squares that can be covered by a set 
of different smaller square \cite{G99}. He has proposed one of the 
most succinct and remarkable Prolog programs to accomplish that task 
efficiently.

\section*{\centerline{Current Work}}
Of late, Alain has also applied himself to the problem of sorting 
interval variables [Alain Colmerauer and No\"elle Bleuzen-Guernalec, 2000]. 
Donald Knuth, one of the world's leaders 
in computer science, has devoted a full volume of his collected 
works to sorting. Alain's involvement in sorting brings a fresh 
new approach to that basic problem. Furthermore, this type of 
sorting has proved to be of paramount importance in scheduling, 
one of the most difficult tasks in Operations Research. Alain's 
approach achieves the complexity of classical sorting and even 
though no large application of the problem is presently known, 
it is not unlikely that it will occur in the future.

Finally, in the past year or so Alain has also renewed his interest 
in an extension of Prolog originally proposed by Michael Maher 
\cite{Mah88}. This extension consists of allowing a user to incorporate 
existential and universal quantifiers to Prolog clauses. Michael 
had written a theoretical account and provided proofs of the 
validity of that approach. Alain's most recent paper demonstrates 
that Maher's ideas are feasible in practice and can be effectively 
used for solving interesting problems [Alain Colmerauer and Dao 
Thi-Bich-Hanh, 2000].

\section*{\centerline{Postlude}}
One can only marvel at the breadth and scope of Alain Colmerauer's 
research. His contributions range from computational linguistics, 
to symbolic manipulation, to language design, to symbolic logic, 
to operations research. This range is matched by in-depth analyses 
of the solutions he has found for complex problems. In the history 
of computer science, the combination of theory and practice present 
in Alain's work has been achieved only rarely. And who knows 
-- Alain may still have a couple of good tricks up his sleeve!
The above paragraph brings to mind a favorite Unix fortune cookie 
saying (those that are used to bid farewell after each session):

\vskip 10pt

\par\noindent\textit{Failure:}
\par\noindent\textit{Work hard to improve}
\par\noindent\textit{Success:}
\par\noindent\textit{You solved the wrong problem}
\par\noindent\textit{Work hard to improve}

\vskip 10pt

The metaphor aptly describes the behavior of a Prolog interpreter 
traversing a search tree while attempting to find all solutions 
to a given program. The metaphor also aptly portrays the tribulations 
and breakthroughs in one's professional life. Perhaps choice, 
genetics, or a combination of both determines the size of our 
own search trees and their number of failure and success nodes. 
Perhaps the number of success nodes could measure our perception 
of a person's achievements. Alain's search tree has proven to 
be quite remarkable! An abundance of success nodes, far removed 
from the root and representing unusual achievements, are present. 
And certainly, his successes have made some of ours possible.\\
Thank you Alain.
\section*{\centerline{Acknowledgements}}
I wish to thank Colette Colmerauer and Krzysztof Apt for the thoughtful 
comments they made on the original manuscript.

\end{document}